\documentclass[11pt]{article}
\usepackage{graphicx}
\usepackage{amssymb}
\usepackage{epstopdf}
\usepackage{amsmath}
\usepackage{natbib}
\usepackage{url}
\usepackage{natbib}
\usepackage{yhmath}
\usepackage{epsfig}
\usepackage{fancyhdr}
\usepackage[hmargin=1in,vmargin=1in]{geometry}
\usepackage{sidecap}
\usepackage[font={small}]{caption}

\begin{document}

\title{Cliffs Benchmarking}
\author{Elena Tolkova \thanks{e.tolkova@gmail.com; elena@nwra.com} \\ NorthWest Research Associates, Redmond, WA, USA}
\maketitle

\pagestyle{fancy}       
\lhead{}

\begin{abstract}
A numerical model for tsunami simulations Cliffs is exercised with the complete set of NTHMP-selected benchmark problems focused on inundation, such as simulating runup of a non-breaking solitary wave onto a sloping beach, runup on a conical island, a lab experiment with a scaled model of Monai Valley, and the 1993 Hokkiado tsunami and inundation of the Okushiri Island.
\end{abstract}

\section{Model Overview}

Cliffs computes tsunami propagation and land inundation under the framework of the non-linear shallow-water theory. Cliffs uses modified finite-difference scheme VTCS-2 \cite[]{titov1995} for numerical integration of the 1D shallow-water equations, dimensional splitting for solving in two spacial dimensions, and a custom land-water interface. Cliffs inundation algorithm is based on a staircase representation of topography and treats a moving shoreline as a moving vertical wall (a moving cliff), which gave the model its name. 

Cliffs performs computations in a single grid given initial deformation of the free surface or the sea floor, and/or initial velocity field, and/or boundary forcing. It also computes boundary time-series input into any number of enclosed grids, to allow further refinement of the solution with one-way nesting. The flow of computations and input/output data types are similar to that in MOST version 4 (not documented; benchmarked for NTHMP in 2011 \cite[]{tolk-nthmp}), which was developed as an adaptation of a curvilinear version of the MOST model \cite[]{curvimost} to spherical coordinate systems arbitrary rotated on the Globe. Cliffs' computational flow has been optimized to focus specifically on geophysical and Cartesian coordinate systems, and to allow easy switch between the coordinate systems, as well as 1D and 2D configurations. Cliffs is coded in Fortran-95 and parallelized using OpenMP. NetCDF format is used for all input and output data files.

Cliffs is developed, documented, and maintained in a GitHub repository\footnote{ \url{https://github.com/Delta-function/cliffs-src}} by E. Tolkova. 
Cliffs code was written in 03/2013-09/2014, and has been occasionally revisited thereafter. 
The model's description is kept current in a detailed User Manual ({\url{http://arxiv.org/abs/1410.0753}}), which comments on deviations from the description given in the model's birth certificate \cite[]{cliffs}. 
Cliffs code is copyrighted under the terms of FreeBSD license. 
\paragraph{The developer's web site} \url{http://elena.tolkova.com/Cliffs.htm} contains links to the model-related resources, including the code repository,  ready-to-try modeling examples, user manual and other related publications. Results of this (and previous) benchmarking with supplementary animations can also be browsed at \url{http://elena.tolkova.com/Cliffs_benchmarking.htm}.

\section{Numerics}

VTCS-2 finite-difference approximation was introduced by VT and CS in 1995 for numerically integrating the non-linear shallow-water equations \cite[]{titov1995}. VTCS-2 difference scheme is described in detail in \citep{titov1995, titov1998}, and presented in a coding-friendly form in \cite[]{cliffs}. Burwell et al (2007) analyzed diffusive and dispersive properties of the VTCS-2 solutions. The present version of Cliffs uses a modification of the VTCS-2 scheme, described in \ref{DS}.

A 1D algorithm can be efficiently applied to solving 2D shallow-water equations using well-known dimensional splitting method \citep{strang, yan, leveque2002}. 
In this way, the VTCS-2 scheme was extended to handle 2D problems in Cartesian coordinates \citep{titov1998}, geophysical spherical coordinates \cite[]{titov97} (the first mentioning of the MOST model), and in an arbitrary orthogonal curvilinear coordinates \cite[]{curvimost}. 

However, the dimensional splitting results in specific sensitivity to the boundary conditions. Cliffs development started with a seemingly minor change to the reflective (vertical wall) boundary condition in VTCS-2/MOST models, which allowed to reduce numerical dissipation on reflection \cite[]{cliffsA}. The modification was extended to include runup computations, by treating a moving shoreline as a moving vertical wall \cite[]{cliffs}. Cliffs inundation algorithm is more compact and approximately 10\% more efficient computationally than the present MOST algorithm.
Below, Cliffs numerics is described in more details, with an emphasize on how Cliffs is different from VTCS-2 / MOST.

\subsection{Difference Scheme}
\label{DS}
The shallow-water equations (SWE) solved in Cliffs are given below in matrix notation in the Cartesian coordinates:
\begin{equation}
W_t=A(W)W_x+B(W)W_y+C(W)
\label{eq1}
\end{equation}
where subscript denotes partial derivatives; $W=\begin{pmatrix} h & u & v \end{pmatrix}^T$ is a vector of state variables; $h$ is height of the water column; $u, \ v$ are particle velocity components in $x$ and $y$ directions;
\begin{equation}
A=-\begin{pmatrix} u & h & 0 \\ g & u & 0 \\ 0 & 0 & u \end{pmatrix}, \ B=-\begin{pmatrix} v & 0 & h \\ 0 & v & 0 \\ g & 0 & v \end{pmatrix}, \ C=\begin{pmatrix} 0 \\  gd_x-\alpha^x \\ gd_y-\alpha^y \end{pmatrix} ;
\label{eqABC}
\end{equation}
$g$ is acceleration due to gravity; $\begin{pmatrix} \alpha^x & \alpha^y \end{pmatrix}$ are the components of acceleration due to friction (Manning friction formulation is used); $d$ is undisturbed water depth, or vertical coordinate of sea bottom measured down from the mean sea level (MSL). Negative values of $d$ correspond to dry land and give the land elevation relative to MSL. \\

Cliffs computes tsunami propagation using a numerical method by Titov and Synolakis (1998), which breaks the original SWE into separate problems of reduced complexity, to be solved sequentially. First, by making use of dimensional splitting, the original 2D problem becomes a sequence of 1D problems for the same state variables, to be solved row-wise and column-wise in an alternate manner: 
\begin{subequations}
\begin{equation}
W_t=A(W)W_x+C_1(W)
\label{eq2A}
\end{equation}
\begin{equation}
W_t=B(W)W_y+C_2(W)
\label{eq2B}
\end{equation}
\label{eq2AB}
\end{subequations}
where $C_1=\begin{pmatrix} 0 & gd_x-\alpha^x & 0 \end{pmatrix}^T$, $C_2=\begin{pmatrix} 0 & 0 & gd_y-\alpha^y \end{pmatrix}^T$.
Next, by transitioning to Riemann invariants 
\[
p=u+2 \sqrt{gh}, \ \ \ q=u-2 \sqrt{gh}
\]
with corresponding eigenvalues $ \lambda_{p,q}=u \pm \sqrt{gh}$, each 1D problem becomes three independent convection problems for a single variable each. The resulting problem set originating with a system \eqref{eq2A} follows:
\begin{subequations}
\begin{equation}
p_t =-(\lambda_1\cdot p_x - g d_x) -\alpha^x
\label{pq1}
\end{equation}
\begin{equation}
q_t=-( \lambda_2\cdot q_x - g d_x) -\alpha^x
\label{pq2}
\end{equation}
\begin{equation}
v_t = - u \cdot v_x
\label{pq3}
\end{equation}
\label{pq_most}
\end{subequations}
\paragraph{Two possible finite difference approximations} for \eqref{pq1} (for \eqref{pq2} analogically) are: 
\begin{equation}
p^{n+1}_j=p^n_j - \Delta t \cdot Q(j+1,j-1)+  \lambda_j  \Delta t^2 \cdot \frac{ \hat{Q}_j- \hat{Q}_{j-1}}{\Delta x_{j-1}+\Delta x_j}-\alpha_j{\Delta t}
\label{solver1}
\end{equation}
or
\begin{equation}
p^{n+1}_j=p^n_j - \frac{\Delta t}{2} \cdot \left( \hat{Q}_{j-1} + \hat{Q}_j  \right)+  \lambda_j  \Delta t^2 \cdot \frac{ \hat{Q}_j- \hat{Q}_{j-1}}{\Delta x_{j-1}+\Delta x_j}-\alpha_j{\Delta t}
\label{solver2}
\end{equation}
where 
\begin{equation}
Q(k,j)=\frac{1}{2} (\lambda_k+\lambda_j) \frac{p_k-p_j}{x_k-x_j}-g\frac{d_k-d_j}{x_k-x_j}, \ \ \ \   \hat{Q}_j=Q(j+1,j)
\label{eqQ}
\end{equation}
with $\Delta x_j=x_{j+1}-x_j$ being a space increment. In the steady state, $Q=0$, which ensures automatic preservation of the steady state. 

The two stencils differ in that 
to approximate a spatial derivative at point $j$, \eqref{solver1} uses a difference across two cells $Q(j-1,j+1)$, whereas \eqref{solver2} uses an average of the left and right one-cell differences $\hat{Q}_{j-1}$ and $\hat{Q}_j$. Since term $Q$ is nonlinear with respect to depth and wave variables 
\[
Q(j-1,j+1) \ne \frac{1}{2}  \left( \hat{Q}_{j-1} + \hat{Q}_j  \right) ,
\]
the two approximations yield noticeably different results for strongly nonlinear problems and/or over large depth variations. 
Both solvers \eqref{solver1} and  \eqref{solver2} are first-order accurate in time and second-order accurate in space for a uniform grid spacing.
Should the spacing $\Delta x_j$ vary, an added numerical error of order $\Delta x_j - \Delta x_{j-1}$ would arise. 
In a basin with constant depth, each scheme is stable under the known limit on the Courant number: $|\lambda| \Delta t/ \Delta x \le 1$. 

Scheme \eqref{solver1} is the original VTCS-2 scheme utilized in MOST; scheme \eqref{solver2} is used in Cliffs. One example where the MOST and Cliffs numerical schemes operate substantially different is given below.

\subsubsection{Dam Break Problem}
\label{DBP}

\begin{SCfigure}
\centering
	\caption{Water profiles 36 s after dam breaking, computed by Cliffs (blue), Cliffs with plugged-in VTCS-2 difference stencil (red), and the analytical solution (black). At $t=0$, 2.5-m-deep water occupied half-space $x>0$.}
		\includegraphics[width=0.7\textwidth]{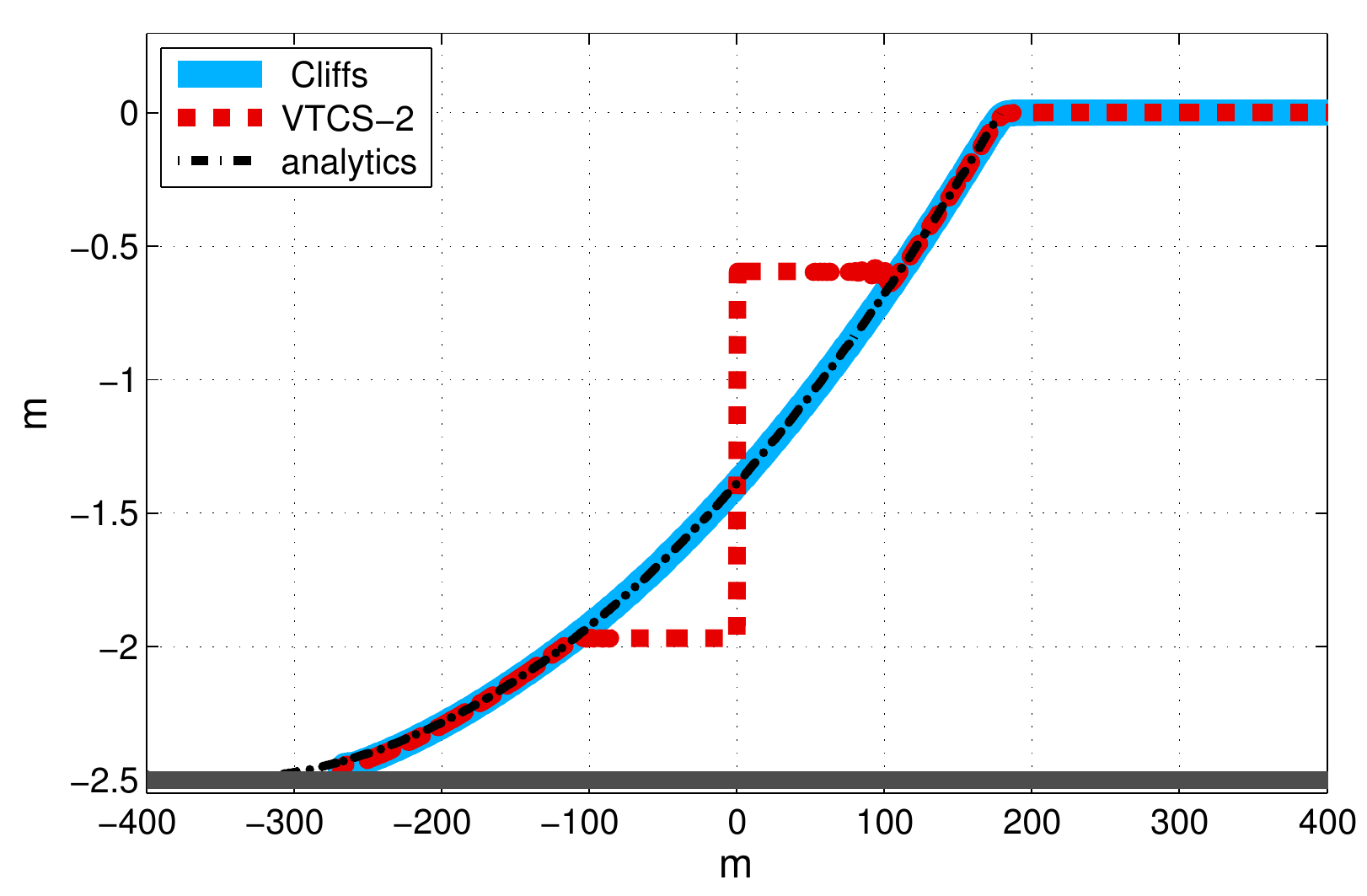}
	\label{dam}
\end{SCfigure} 

Correct application of the VTCS-2 stencil requires the characteristics of the same family not to change their direction in three successive nodes - a condition which cannot incorporate rarefaction waves \cite[]{leveque2002}. Consequently, VTCS-2 scheme fails to correctly simulate a classical Dam Break problem\footnote{Dam break problem is a frequent textbook example of the shallow-water flow with known analytical solution. In spite of its popularity, this problem is omitted in PMEL-135
Report setting benchmarking criteria \cite[]{synolak2007}, and consequently, it is not included with the NTHMP's set of inundation benchmarks.} \citep{stoker, leveque2002}.
As seen in Fig. \ref{dam}, a few modifications to VTCS-2 difference scheme implemented in Cliffs has enabled the model to handle the Dam Break problem. Figure \ref{dam} displays simulated profiles of water volume 36 s after instantaneous removal of a 2.5-m-high dam  vs. the analytical solution. At $t=0$, 2.5-m-deep water occupied half-space $x>0$. The solutions to the problem are computed by Cliffs with its modified VTCS-2 solver, and by Cliffs with plugged-in original VTCS-2. The moving shoreline algorithm is that of Cliffs model in both cases. For this problem, Cliffs solver yields results coinciding with the theoretical solution, whereas VTCS-2 / MOST solver computes an unrealistic discontinuous waveform.

\subsection{Land-water interface: vertical wall}

According to VTCS-2 algorithm, the sea-going Riemann invariant  in the last wet node next to a reflective boundary (a vertical wall) is assigned a value opposite to that of the wall-going invariant (Titov and Synolakis, 1995, 1998). 
This treatment sets the reflective wall immediately next to the last wet node. However, as discussed below, this treatment does not mix well with the splitting technique.

\begin{SCfigure}
\centering
	\caption{Black dots - water, circles - land. Land-water boundary seen by MOST in x-direction (black line) and in y-direction (gray line) are shown on north-west channel boundary. Land-water boundary seen by Cliffs in either direction is shown on south-east channel boundary.}
		\includegraphics[width=0.3\textwidth]{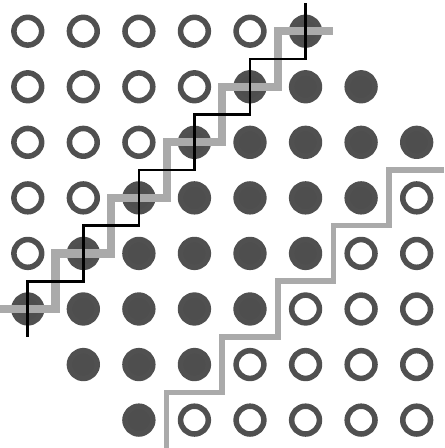}
	\label{dots}
\end{SCfigure}

Unless a straight boundary coincides with either x or y axis, it can not be treated by a splitted scheme as straight, but rather as a step-like. Consider, for example, a diagonal channel shown in Figure \ref{dots}. When the splitted VTCS-2 solver performs computations in x-direction, it ``sees" this boundary as a set of steps whose vertical segments are aligned with the edge wet nodes, as shown on the north-west channel boundary with a black line. When computations are performed in y-direction, the solver ``sees" a different set of steps. Now the horizontal segments of the steps are aligned with the edge nodes, as shown by a gray line on the north-west boundary. This displacement of the reflective boundary within a splitted cycle results in increased dissipation, occasionally causing unrealistically low later waves (reflected from coastlines) in simulations of historical tsunami events with MOST.

In Cliffs, the reflective boundary conditions are formulated by introducing a mirror ghost node coinciding with the first dry node. This treatment sets the reflecting wall exactly in the middle between the neighboring wet and dry nodes, where it remains steady for the entire splitted cycle (see Fig. \ref{dots}). This modification has resulted in better representation of the later waves in simulations of the real-world tsunamis \citep{cliffs}.

\subsection{Land-water interface: moving shoreline} 

Unlike the VTCS-2 approach, Cliffs boundary conditions are obtained without making any assumptions about the direction of the characteristics, which allows to apply them on a moving boundary as well.
Hence Cliffs inundation algorithm is based on a staircase representation of topography, and treats a vertical interface between wet and dry cells (a shoreline) as a vertical wall. Reflective boundary conditions applied on an instant shoreline are formulated using a mirror ghost node coinciding with a shoreline dry node. 

To enable wetting/drying, a common procedure of comparing the flow depth with an empirical threshold $h_{min}$ is used. 
The wet area expands when a flooding depth in a shoreline wet node exceeds a threshold (runup), and shrinks when a flow depth in a cell is below the threshold (rundown). On the wet area expansion, dry cells to be flooded are pre-filled with water to $h_{min}$ depth. Hence the runup occurs on an inserted cushion $h_{min}$ high which is removed on rundown. 

\section{Testing Cliffs with NTHMP Benchmark Problems}

Cliffs is seeking to join a list of ten (on 2016) tsunami models endorsed by the National Tsunami Hazard Mitigation Program (NTHMP) after passing the inundation benchmarks, which were selected according to NOAA's standards and criteria \citep{synolak2007,nthmp}.  
Detailed descriptions of the benchmarks, as well as topography and laboratory or survey data when applicable, can be found in a repository of benchmark problems \url{https:// github.com/ rjleveque/ nthmp-benchmark-problems} for NTHMP, or in the NOAA/PMEL repository \url{http:// nctr.pmel.noaa.gov/ benchmark/}. Cliffs performance with these benchmark problems (BPs) is presented below. Wherever applicable, the results are displayed and the errors are evaluated with the NTHMP provided scripts. Current Cliffs benchmarking results with complementary animations can also be found at \url{http://elena.tolkova.com/Cliffs_benchmarking.htm}.

\subsection{BP1 - Solitary wave on a simple beach (nonbreaking - analytic)}

\begin{figure}[ht]
	\resizebox{1.1\textwidth}{!} 
		{\includegraphics{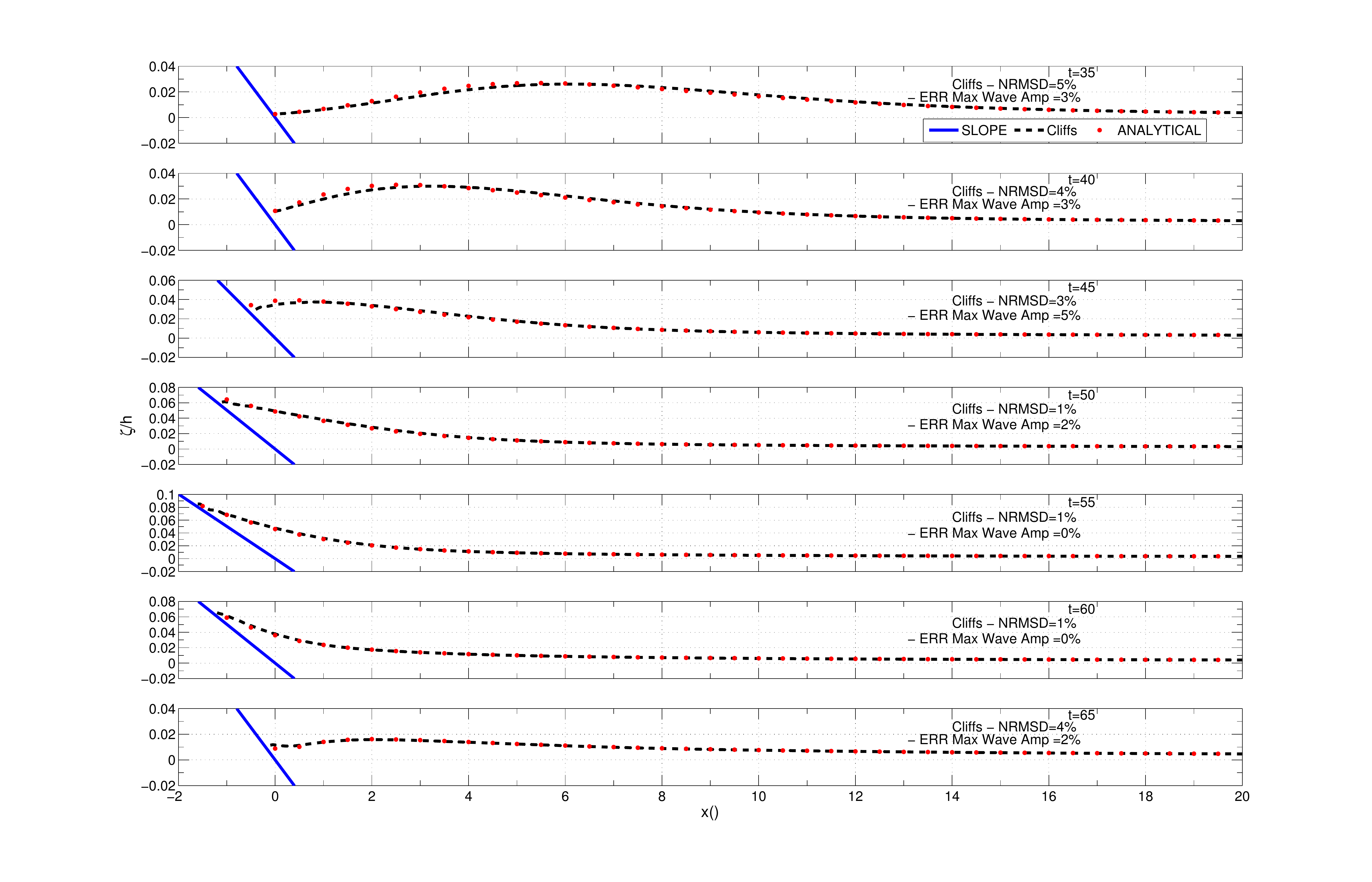}}
	\caption{Water surface profiles for an initial solitary wave $0.0185 d$ high climbing up a 1:19.85 beach at $t (g/d)^{1/2}=35,40,45,50,55,60,65$ (shown in a plot), black - Cliffs numerical solution, red dots - analytical solution. Plotted with NTHMP provided script.  }
	\label{Sprf185}
\end{figure}
\begin{figure}[ht]
	\resizebox{\textwidth}{!} 
		{\includegraphics{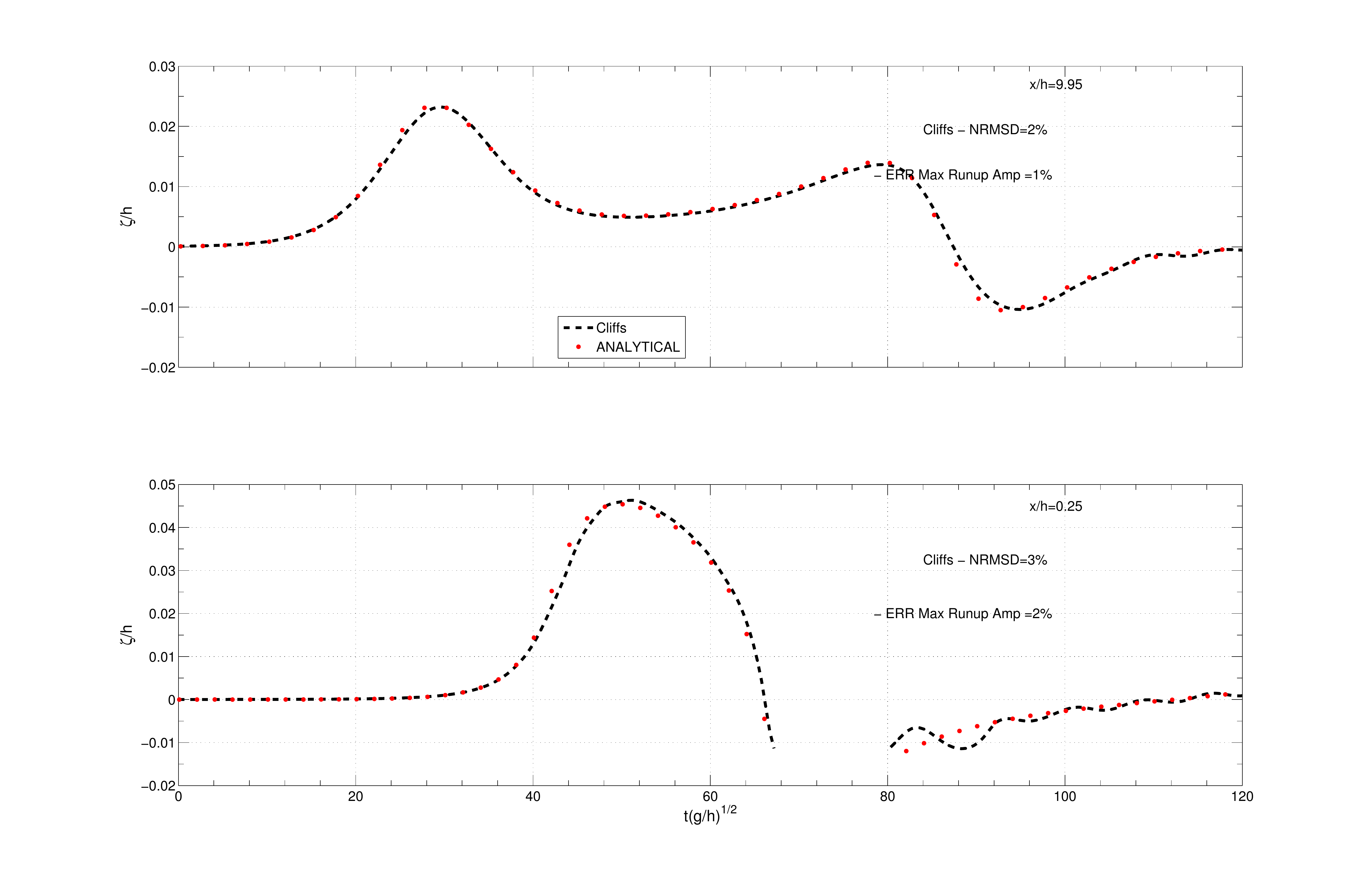}}
	\caption{Water surface time histories at locations $x=9.75 d$ ($0.517 d$ deep) and $x=0.25 d$ ($0.014 d$ deep) in a solitary wave of $0.0185 d$ initial height:  Cliffs numerical solution  (black dashed) vs analytical solution (red dots). The solutions are interrupted in the second location,  when the location temporally dries out. Plotted with NTHMP provided script. }
	\label{Stmsers}
\end{figure}

In this BP,  a non-breaking solitary wave of initial height $H=0.0185d$ over depth $d$ normally approaches and climbs onto a plane sloping beach. 
The Objective for this BP is to model the surface elevation in space and time within 5\% of
the calculated value from the analytical solution.

The geometry of the beach and the wave-profile are described in many articles (\citep{synolak1987, titov1995, synolak2007}) as well as in the Internet repositories mentioned above. The 1D bathymetry consists of a flat segment of depth $d$ connected to a beach with a slope 1:19.85. The x coordinate increases seaward, $x=0$ is the initial shoreline position, and the toe of the beach is located at $x=19.85d$.  
The runup simulation of a non-breaking solitary wave  with $H/d=0.0185$ was performed with Cliffs using a 384-node grid, which encompassed a $45d$-long segment of constant depth $d$ connected to $50d$-long 1:19.85 slope.  In simulations, the depth of the flat part of the basin was $d=1$ m. The grid spacing was set to 1 m (depth) over the flat segment, and then varied as $\sqrt{d}$, but not less than $0.1d$. Time increment $\Delta t=0.03$ s yields Courant number $\Delta t {\sqrt{gd}} / \Delta x =0.1$, thus providing for emulating physical dispersion with numerical dispersion \citep{burwell, tolk-nthmp}. Friction coefficient was set to zero. Depth threshold $h_{min}$ was set to 2 mm. Results with dimension of length are expressed in units of $d$, time is expressed in units of $\tau=\sqrt{d/g}$. 

Figure \ref{Sprf185} shows water surface profiles computed with Cliffs vs. an analytical solution provided by NTHMP. Figure \ref{Stmsers} shows surface elevation time histories computed with Cliffs vs. the analytical solution. Cliffs solution approximated the analytical solution well within  the limits of the no-more-than-5\% error. Namely, the mean normalized standard deviation and max wave amplitude errors, as defined by the NTHMP \cite[]{nthmp}, are 3\% and 2\% respectively for the computed surface profiles, 2\% and 1\% for the surface elevation time-history at the first location, and 3\% and 2\%  for the surface elevation time-history at the second location. 

Maximal computed runup height is $0.087d$.

\subsection{BP4 - Solitary wave on a simple beach (nonbreaking - lab)}

\begin{figure}[ht]
	\resizebox{1.1\textwidth}{!} 
		{\includegraphics{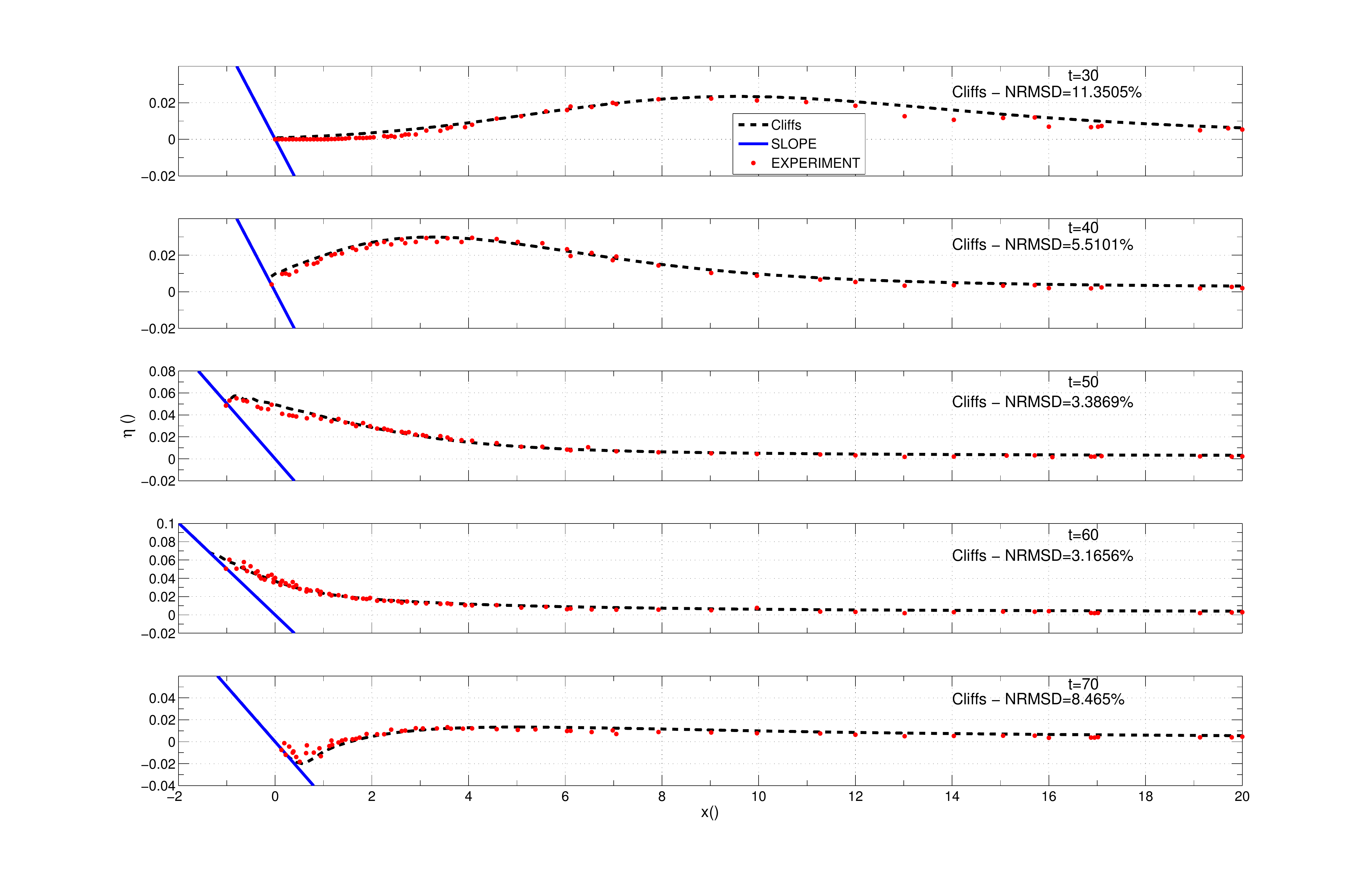}}
	\caption{Water surface profiles for an initial solitary wave $0.0185 d$ high climbing up a 1:19.85 beach at $t (g/d)^{1/2}=30,40,50,60,70$ (shown in a plot), black - Cliffs numerical solution, red dots - lab data. Plotted with NTHMP provided script.  }
	\label{Slab185}
\end{figure}

The Objective of this BP is to approximate lab measurements of a solitary wave attacking a plane sloping beach, with the same geometry as in BP1. Simulation set-up was the same as for BP1, but friction with Manning roughness coefficient $n=0.015$ was included in the simulation aiming to reproduce the lab experiment.
Figure \ref{Slab185} shows water surface profiles computed with Cliffs vs. lab data provided by NTHMP. 
Cliffs' errors -- mean normalized standard deviation (N.St.D.) and max wave amplitude error -- are 6\% and 4\% respectively. For comparison, among all NTHMP-approved tsunami inundation models, the mean N.St.D. errors range is 7-11\%, and mean MaxWave errors range is 2-10\% \cite[]{nthmp}.
Maximal computed runup height is $0.077d$ (lower runup value is due to friction).

\clearpage

\subsection{BP6: Solitary wave on a Conical Island}
\label{sec_island}

This BP requires to numerically simulate a wave-tank experiment where a plane solitary wave attacks and inundates a conically shaped island \citep{briggs, liu1995}. A model of the conical island with 1:4 slope and 7.2 m diameter at the base was constructed near the center of a flat $29.3 \times 30$ m basin filled with water $d=32$ cm deep. In three separate experiments, three different incident solitary waves were generated: $0.05d$, $0.1d$, and $0.2d$ high, labeled cases A, B, and C correspondently.
In each case, time histories at several locations around the island, and the angular distribution of runup were recorded. 
The Objective for this BP is to predict the runup measurements around the island with no more that 20\% mean errors as calculated by the NTHMP provided scripts.

The problem was simulated using two nested grids. The outer grid at $30$ cm spacing enclosed the entire basin area starting at the wave-maker. Simulations were forced through the western boundary, for the duration of the direct pulse. Boundary velocity was computed according to the paddle trajectories and complemented with the surface elevation as in a purely incident wave.
Run-up onto the island was simulated with a finer 10 m x 10 m grid spaced at $5$ cm; the roughness coefficient was $n=0.015$;  $h_{min}=5$ mm. The inundated area was obtained using Cliffs'  ``maxwave" output. The inundation boundary was drawn through interfaces between always dry nodes (NaNs in the maxwave output) and the nodes which get wet at any time during the simulation (have valid maxwave values).

\begin{figure}[ht]
	\resizebox{\textwidth}{!} %
		{\includegraphics{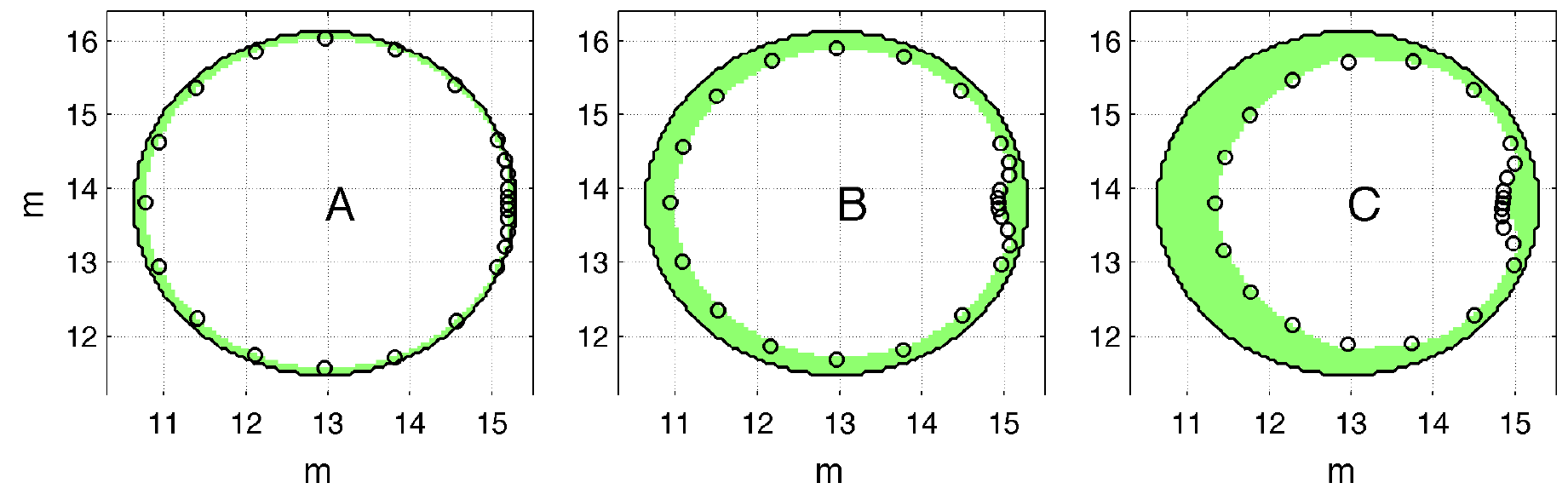}} 
	\caption{Inundation around the island in cases A, B, and C: computed with Cliffs (green), laboratory measurements (black circles), original coastline (black); wave attacks from the west. }
	\label{conical1}
\end{figure}

\begin{figure}[ht]
	\resizebox{1.1\textwidth}{!} 
		{\includegraphics{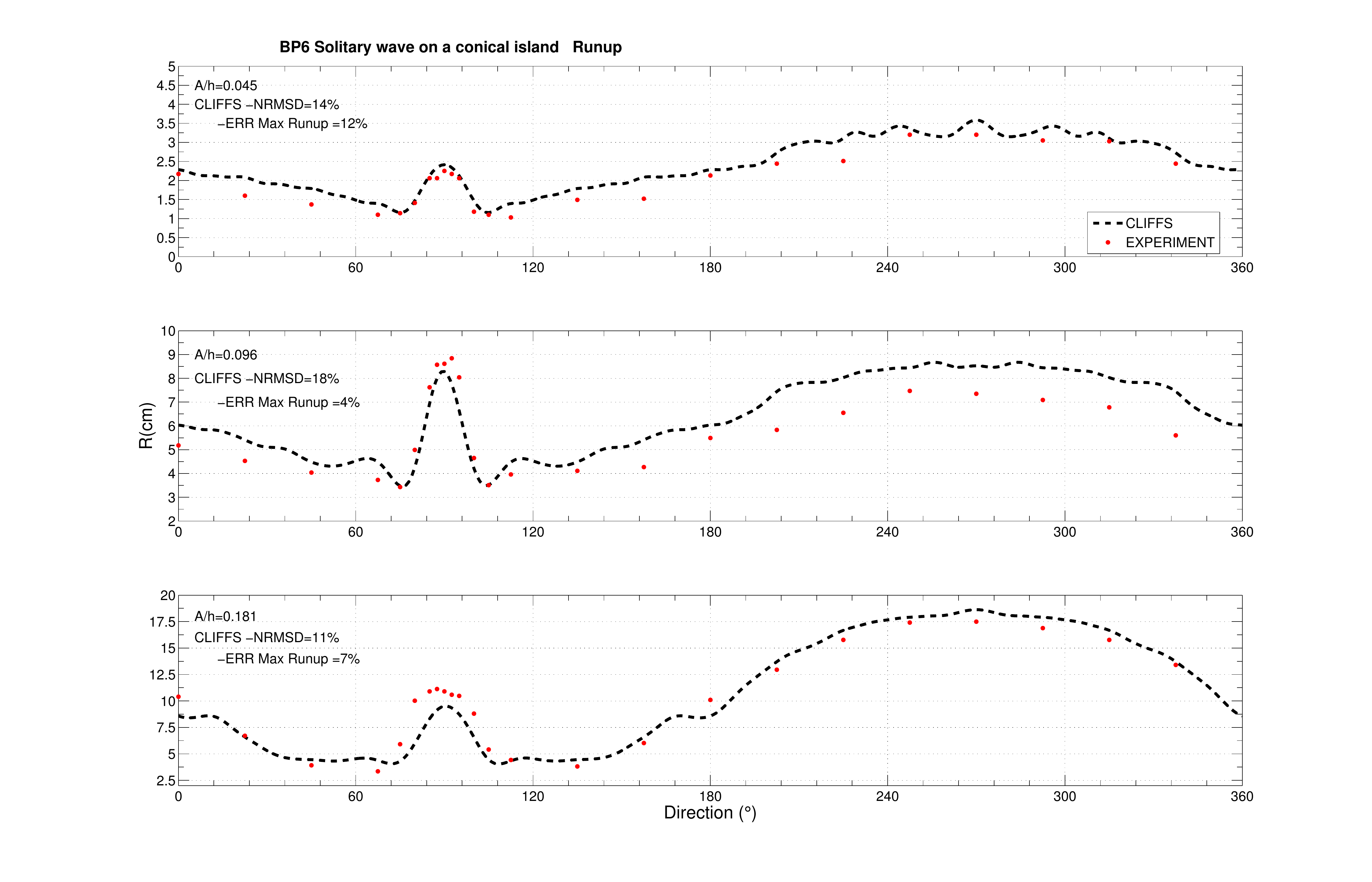}}
	\caption{Runup distribution around the island in cases A, B, and C: computed with Cliffs (black) vs. laboratory measurements (red).  Plotted with NTHMP provided script. }
	\label{conical2}
\end{figure}

\begin{figure}[ht]
	\resizebox{1.1\textwidth}{!} 
		{\includegraphics{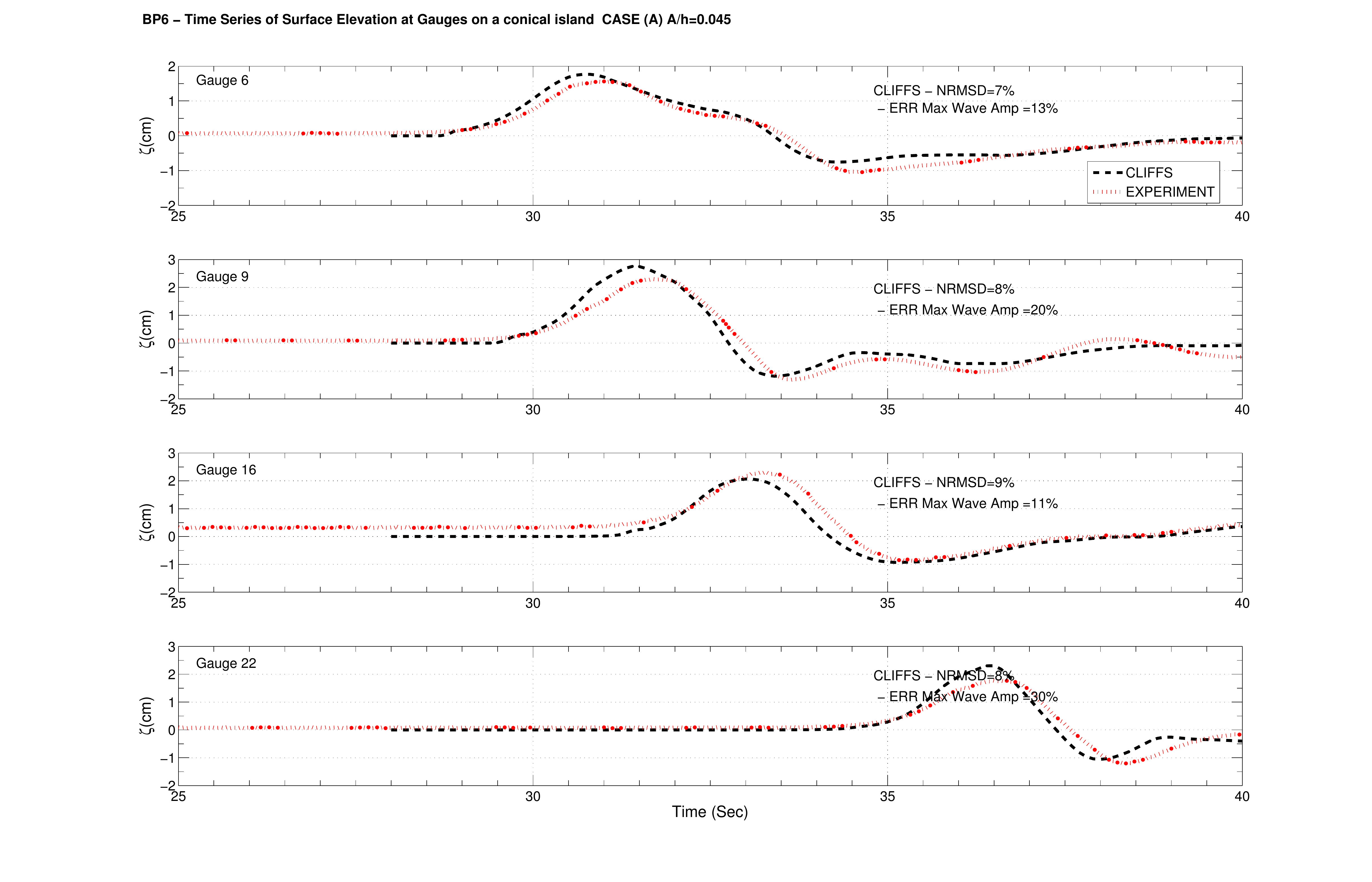}}
	\caption{Time histories at the gages around the island, recorded (red) and computed with Cliffs (black). Case A.  Plotted with NTHMP provided script. }
	\label{conical3A}
\end{figure}

\begin{figure}[ht]
	\resizebox{1.1\textwidth}{!}
		{\includegraphics{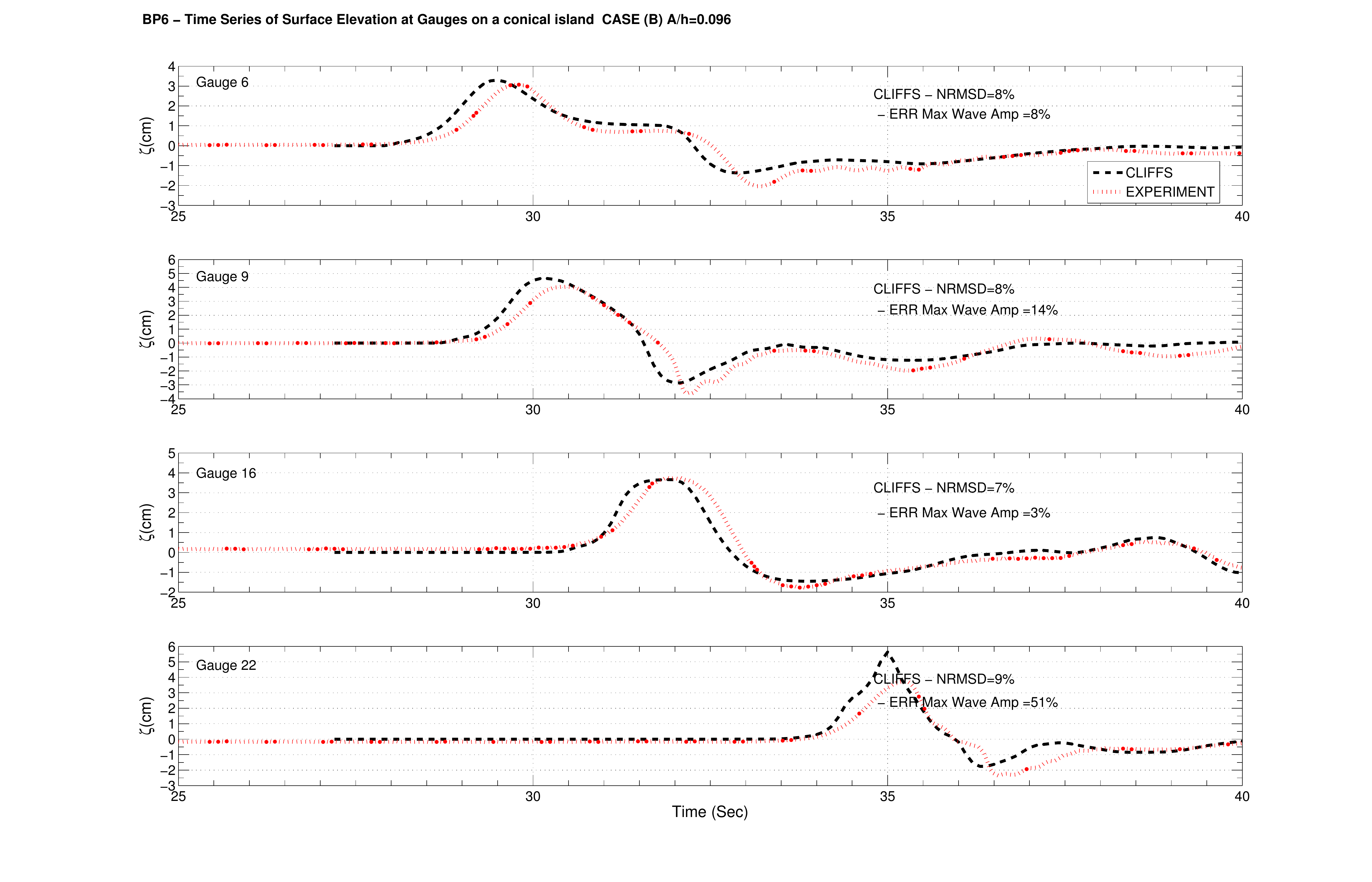}}
	\caption{Time histories at the gages around the island, recorded (red) and computed with Cliffs (black). Case B.  Plotted with NTHMP provided script. }
	\label{conical3B}
\end{figure}

\begin{figure}[ht]
	\resizebox{1.1\textwidth}{!} 
		{\includegraphics{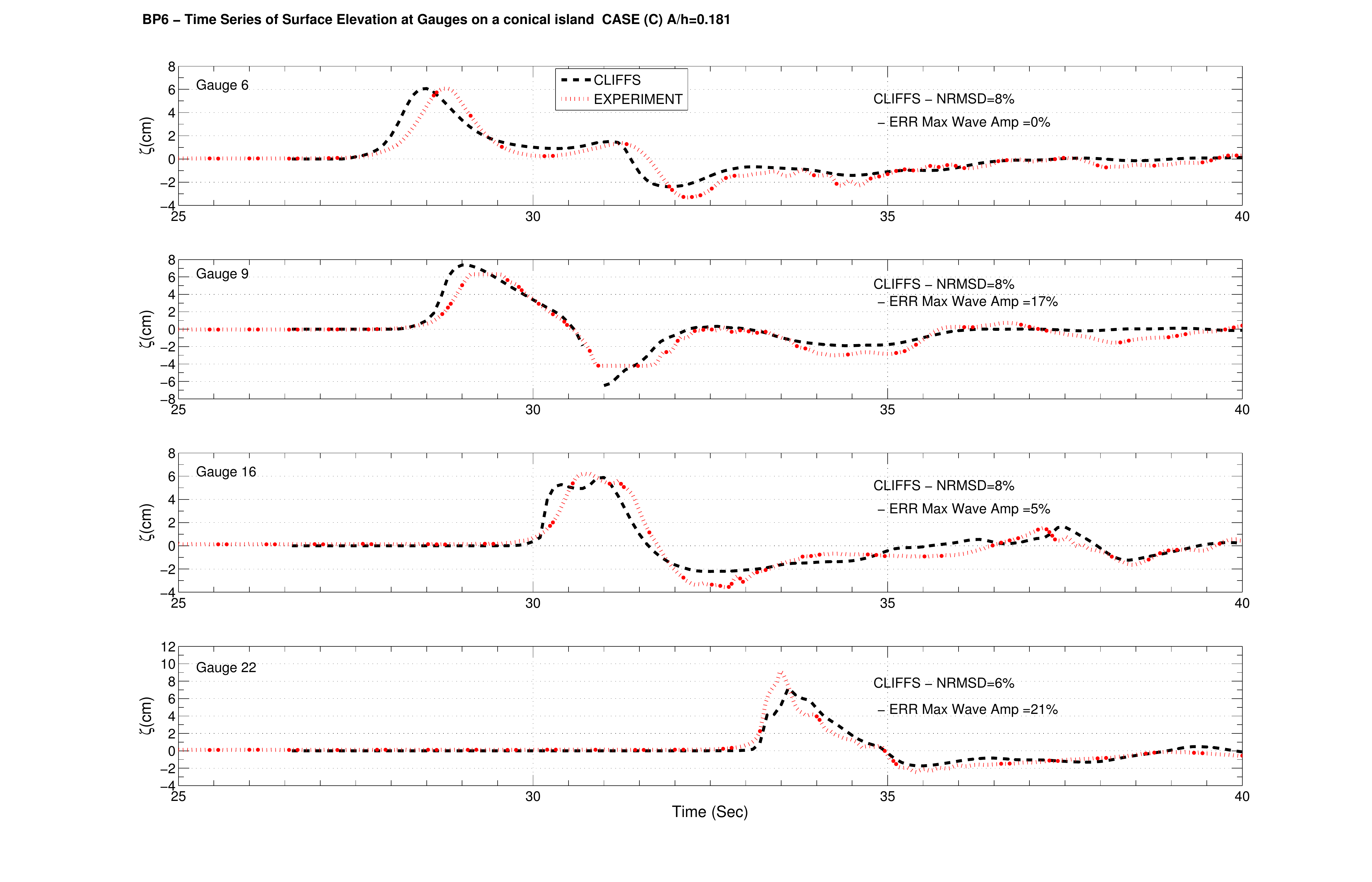}}
	\caption{Time histories at the gages around the island, recorded (red) and computed with Cliffs (black). Case C.  Plotted with NTHMP provided script. }
	\label{conical3C}
\end{figure}

The computed inundated area around the island and the runup height distribution in each of the three cases vs. measurements are shown in Figures \ref{conical1} - \ref{conical2}. Simulated time histories at four gages closest to the island (gage 6 at depth 31.7 cm, gage 9 at 8.2 cm, gage 16 at 7.9 cm, and gage 22 on the lee side at depth 8.3 cm)  vs. laboratory measurements for the three cases are shown in Figures \ref{conical3A}-\ref{conical3C}. As seen in Table \ref{tblcon}, all errors are below the 20\% threshold.


\begin{table}
\centering
\begin{tabular}{c|ccc|ccc}
Errors& & Runup & & & Gauges & \\
& A & B & C & A & B & C \\ \hline 
N.St.D., \%&14 &18 &11 & 8&8 &7.5 \\
Max., \%& 12&4 &7 & 18.5& 19& 11\\
\end{tabular}
\caption{Relative errors of Cliffs computations of the runup heights around the island, and of the water level time-histories at the gages.}
\label{tblcon}
\end{table}

\clearpage

\subsection{BP7: Runup onto a lab model of the Monai Valley}
\begin{figure}[ht]
\centering
	\resizebox{0.7\textwidth}{!} 
		{\includegraphics{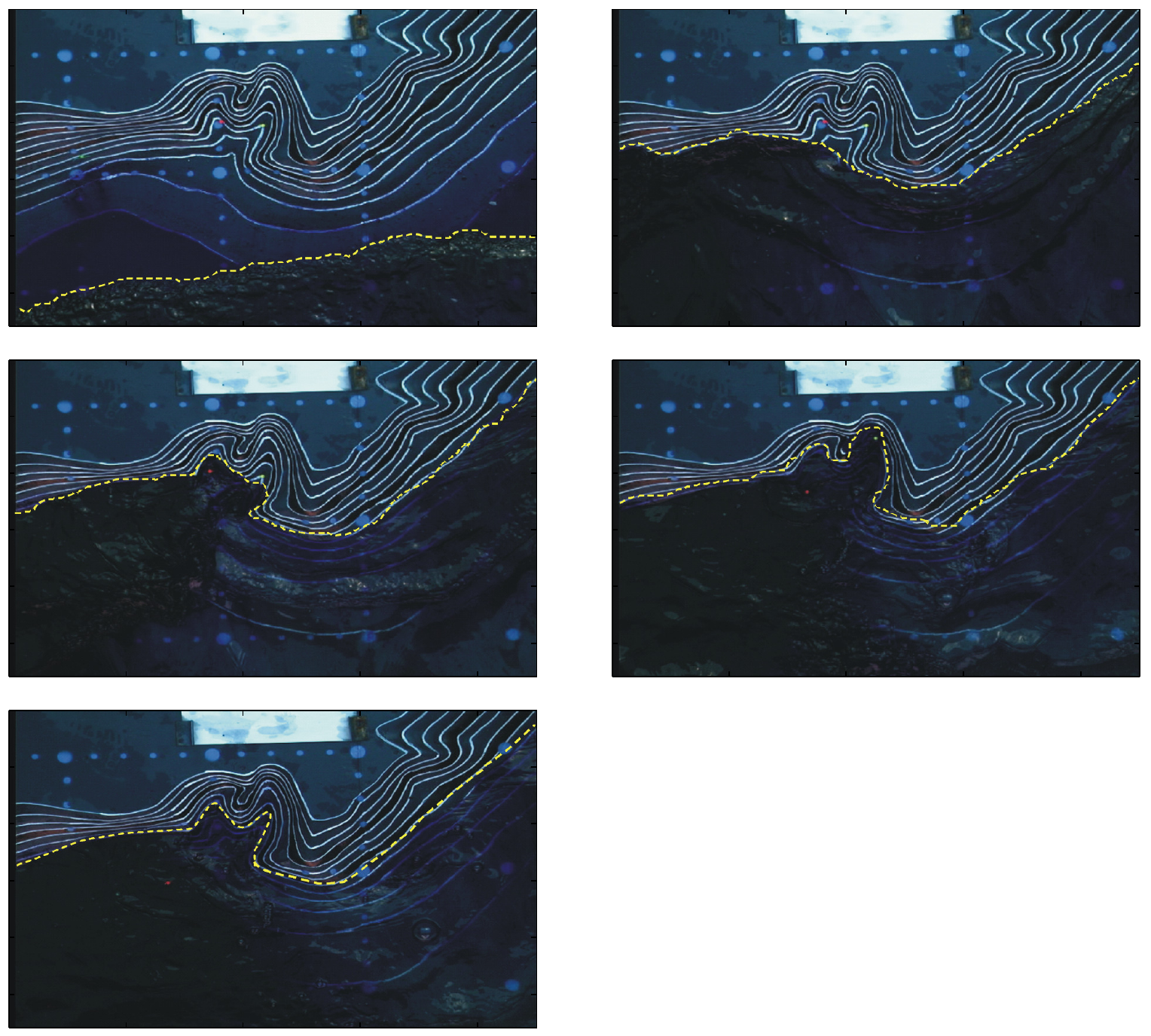}} 
	\caption{Five frames extracted from the video record of the lab experiment with 0.5 sec interval}
	\label{monai2}
\end{figure}
\begin{figure}[ht]
\centering
	\resizebox{0.7\textwidth}{!} 
		{\includegraphics{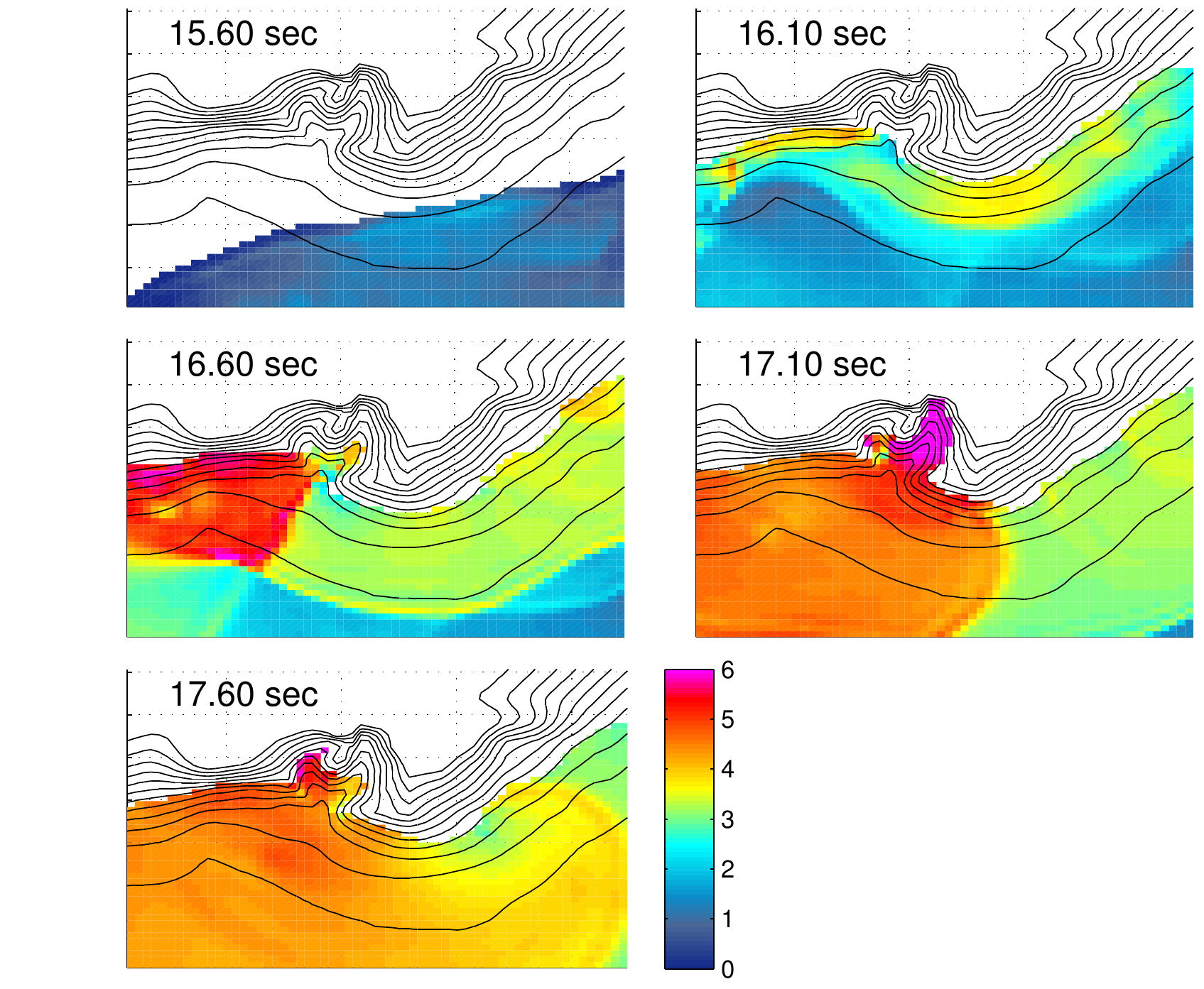}}
	\caption{Five snapshots of simulated water height distribution with 0.5 s interval. }
	\label{monai3}
\end{figure}

\begin{SCfigure}
	\caption{Modeled (green) and recorded (black) water level time histories at the gages 5, 7, 9 (top to bottom).}
		\includegraphics[width=0.5\textwidth]{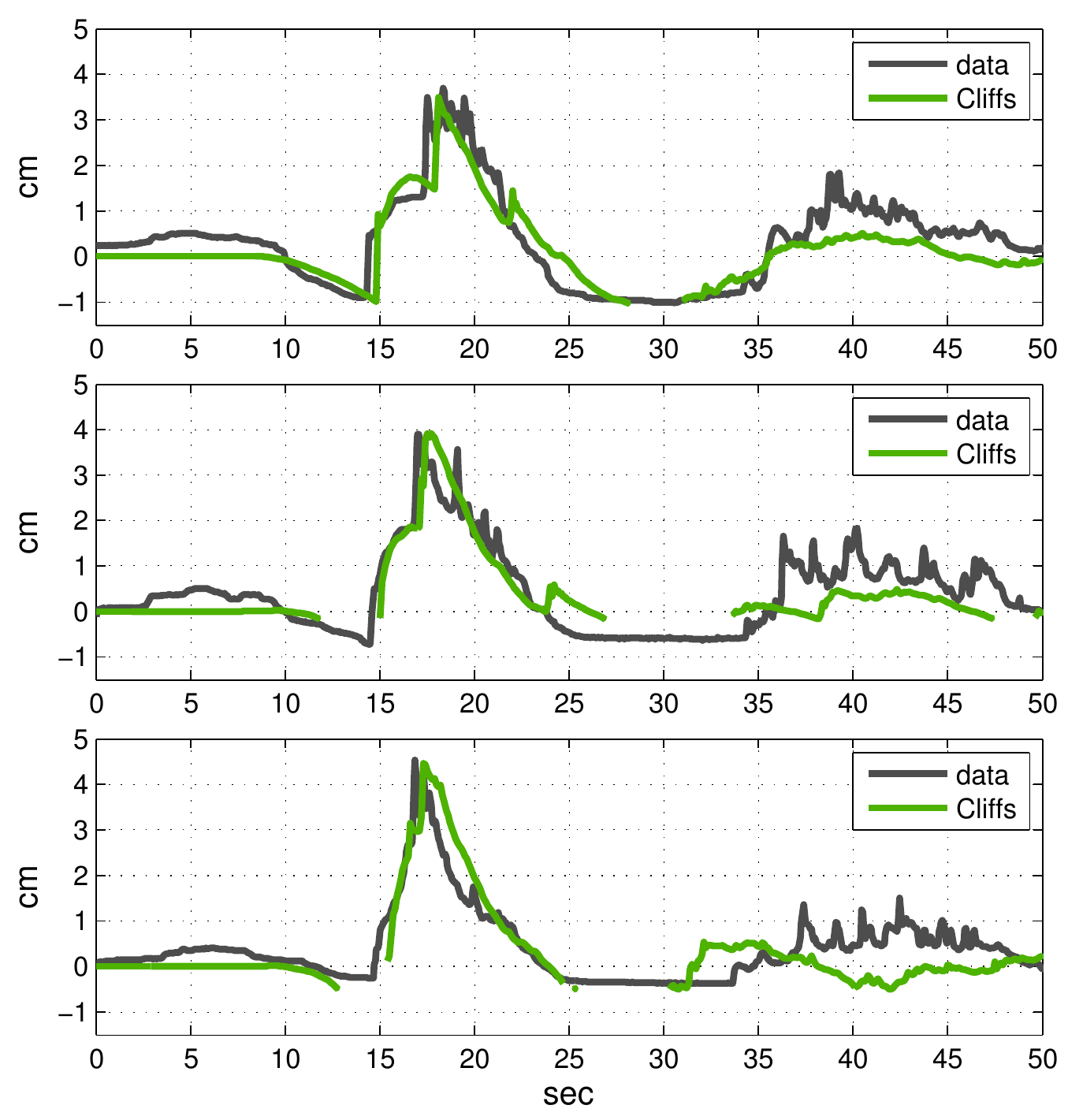} 
	\label{monai1}
\end{SCfigure}

The 1993 Hokkaido tsunami (see BP9) caused extreme runup at the tip of a narrow gully at Monai, Okushiri island, Japan.  
This event was reproduced in a wave tank experiment with a 1:400 scale version of the Monai valley  \citep{liu2008, synolak2007}. 
The Objective of this BP is demonstrating a tsunami model's ability to capture shoreline dynamics with an extreme runup and rundown over a complex topography. 

The domain of computations represents a $5.5 \times 3.4$ m portion of the tank next to the shoreline. Water level dynamics in this region were recorded on video and at three gages along the shoreline. Five frames extracted from the video record of the lab experiment with 0.5 sec interval \cite[]{nicolsky2011} are shown in Figure \ref{monai2}. The first frame occurred approximately at 15.3 s of the lab event time.  

The lab experiment has been modeled with a provided grid at 1.4 cm spacing, friction coefficient $n=0.012$, $h_{min}=1$ mm. Simulations were forced with an incident wave through the sea-side boundary of the domain prescribed for the first 23 s.
Five sea surface snapshots 0.5 s apart are shown in Figure \ref{monai3}. The first snapshot was taken at 15.6 s. By visual comparison, the simulations agree closely with the recorded shoreline position in space and in time. Close fit in the direct wave is also observed between modeled and recorded water level time histories at the gages, in Figure \ref{monai1}. Maximal computed runup height is 8.6 cm.

\clearpage

\subsection{BP9: Okushiri Island tsunami (field)}

\begin{figure}[ht]
\centering
	\resizebox{0.9\textwidth}{!} 
		{\includegraphics{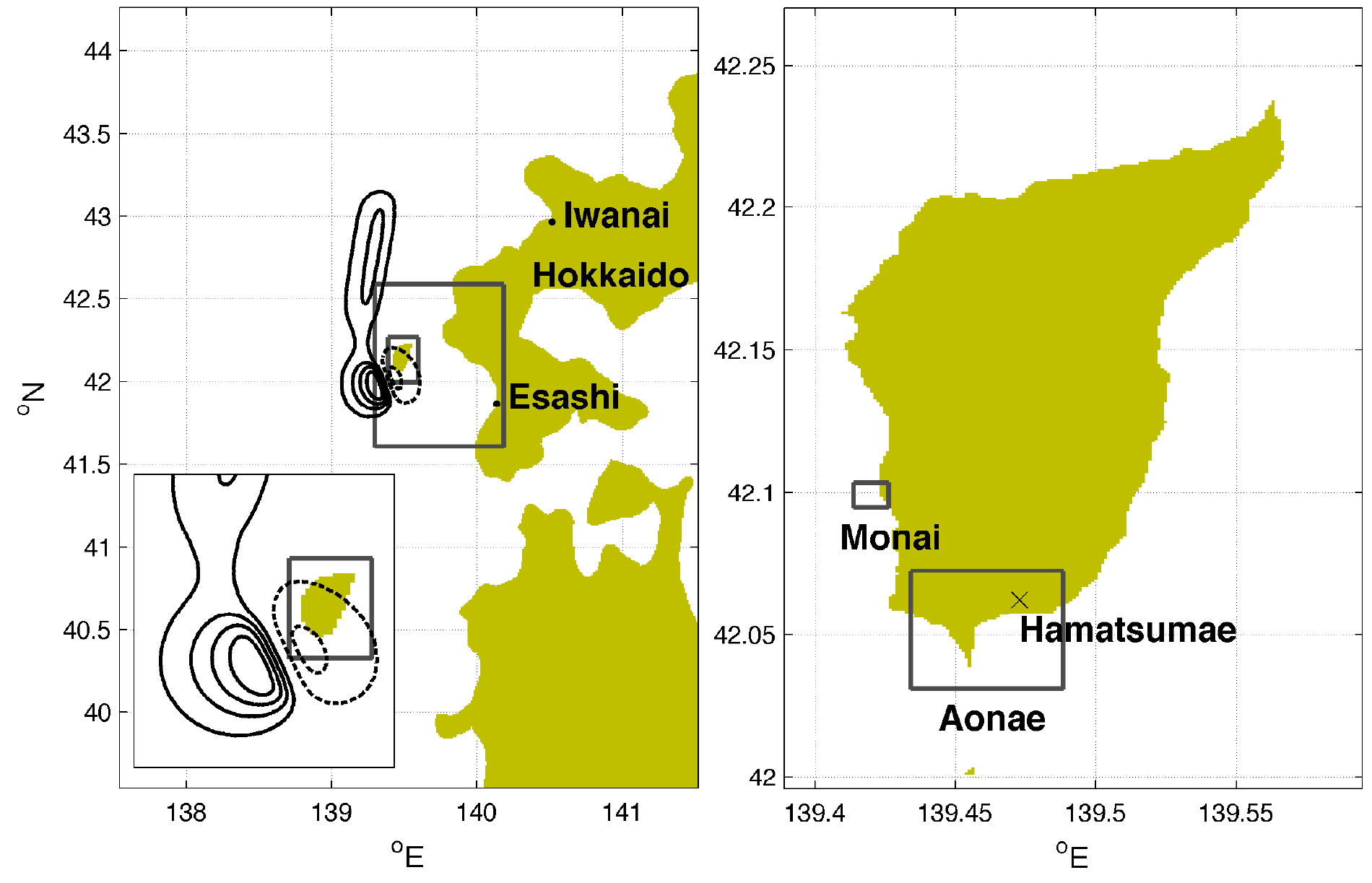}}
	\caption{Left: computational domain used to simulate the 1993 Hokkaido tsunami with contours of the two nested grids; initial sea surface deformation with contour lines at -1, -0.5, 1, 2, 3, 4 m levels, subsidence contours are shown with dashed lines, uplift with solid lines. The deformation area near Okushiri is zoomed-in in the bottom left corner. Right: 3-d nesting level grid around Okushiri island with contours of the 4-th level grids around Monai and Aonae.}
	\label{OKsetup}
\end{figure}

\begin{SCfigure}
	\caption{NTHMP's reference points (blue circles), field measurements (red dots), and boxes to describe modeled runup in terms of minimal, maximal and mean heights.}
		\includegraphics[width=0.5\textwidth]{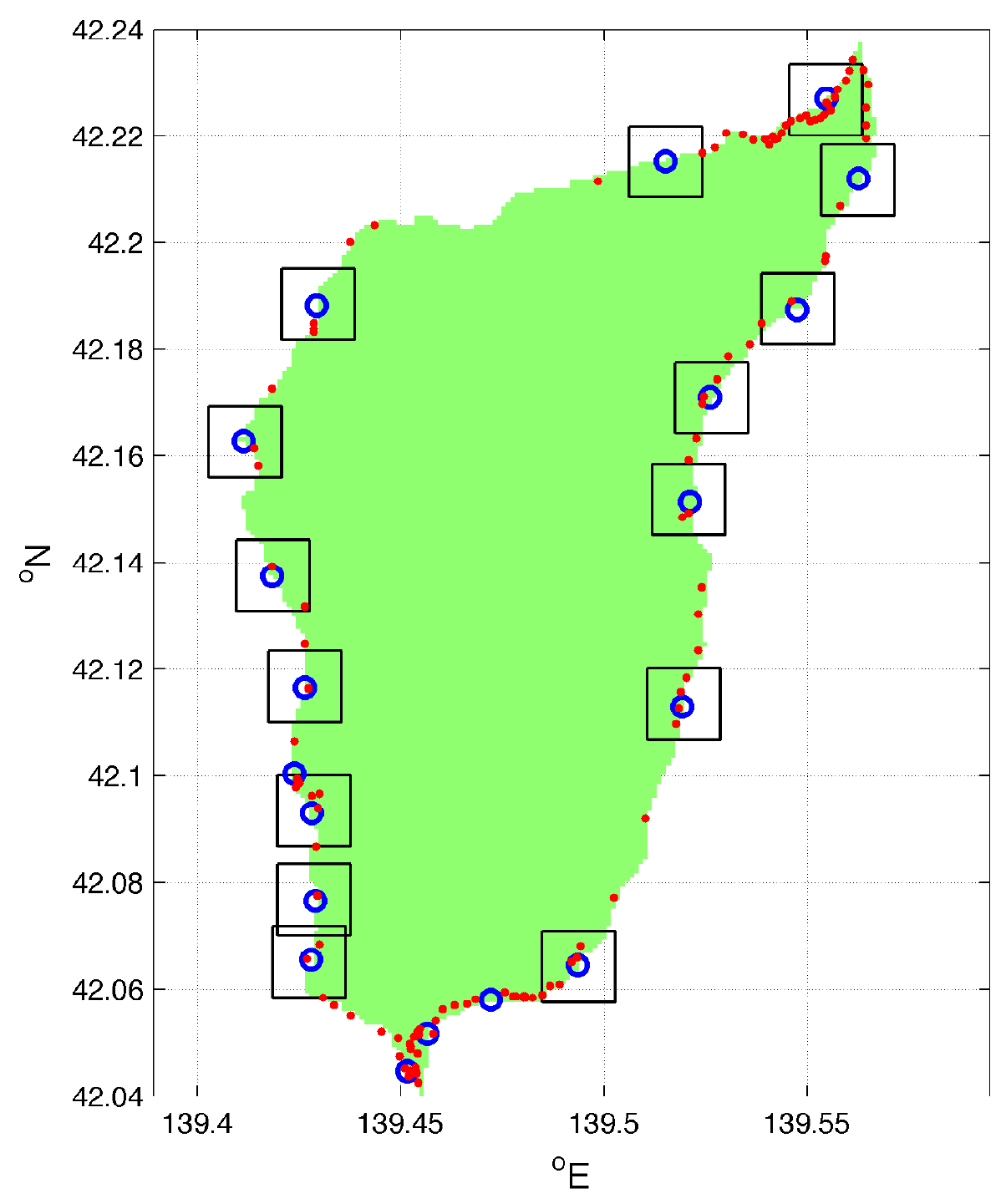} 
	\label{OKsurvey}
\end{SCfigure}

On July 12, 1993, the Mw 7.8 earthquake west of Okushiri island, Japan, generated a tsunami that has become a test case for tsunami modeling efforts \citep{takahashi, synolak2007, nthmp}. Detailed runup measurements around Okushiri island were conducted by the Hokkaido tsunami survey group which reported up to 31.7 m runup near Monai village. Also, high-resolution bathymetric surveys were performed before and after the EQ, which allowed to evaluate the deformation of the ocean bottom due to the quake. The conventional practice in modeling geophysical tsunamis is to assume that the tsunami originates with the initial deformation of the sea surface following that of the ocean bottom.

The computational domain used to simulate the 1993 Okushiri tsunami and the initial sea surface deformation are shown in the left pane in Figure \ref{OKsetup}. The computations at a resolution of 30 sec of the Great arc were refined with nested grids spaced at 10 arc-sec, 3 arc-sec (circling Okushiri island), 15 m (enclosing Aonae area), and 6 m (enclosing Monai valley). The grids contours are pictured in Figure \ref{OKsetup}. 
In the two outer level grids, a vertical wall was imposed in 1 m deep water.
In the 3 arc-sec grid around Okushiri, waves higher than $h_{min}=0.5$ m were permitted to inundate the next dry cell, that is, to advance by another 90 m inland. In the two finest grids, a wave was permitted to inundate once it was higher than $h_{min}=0.1$ m. Friction coefficient was set to $n=0.025$ in the two outer grids, $n=0.04$ in Okushiri-sized grid, and $n=0.03$ in the two finest grids.

This BP has several objectives. The fist Objective is to model the inundation around Okushiri Island and match field measurements of runup heights with no less than 20\% accuracy. In the NTHMP-provided script, runup error is evaluated by a sophisticated comparison between computed and measured sets of minimal, maximal, and mean runup values in unspecified surroundings of prescribed reference points. Figure \ref{OKsurvey} shows locations of the NTHMP's reference points (blue circles) and field measurements (red dots). Considering that 
(1) the reference point surroundings should better contain at least one field measurement, but preferably a few; and 
(2) uncertainty in geo referencing is about 0.011 deg = 1.2 km;
the range of computed runup values was taken from a box with 1.5 km side (as shown in the figure) around each reference point in the 3 arc-s Okushiri grid; and from a box with 120 m side in both Aonae and Monai grids. As evaluated by NTHMP's script, Cliffs reproduced the observed runup with an error of 15\%.  

\begin{SCfigure}
	\caption{Runup around Aonae peninsular computed with Cliffs (black thin vertical bars) and the survey data (orange thick bars), per observation location.}
		\includegraphics[width=0.7\textwidth]{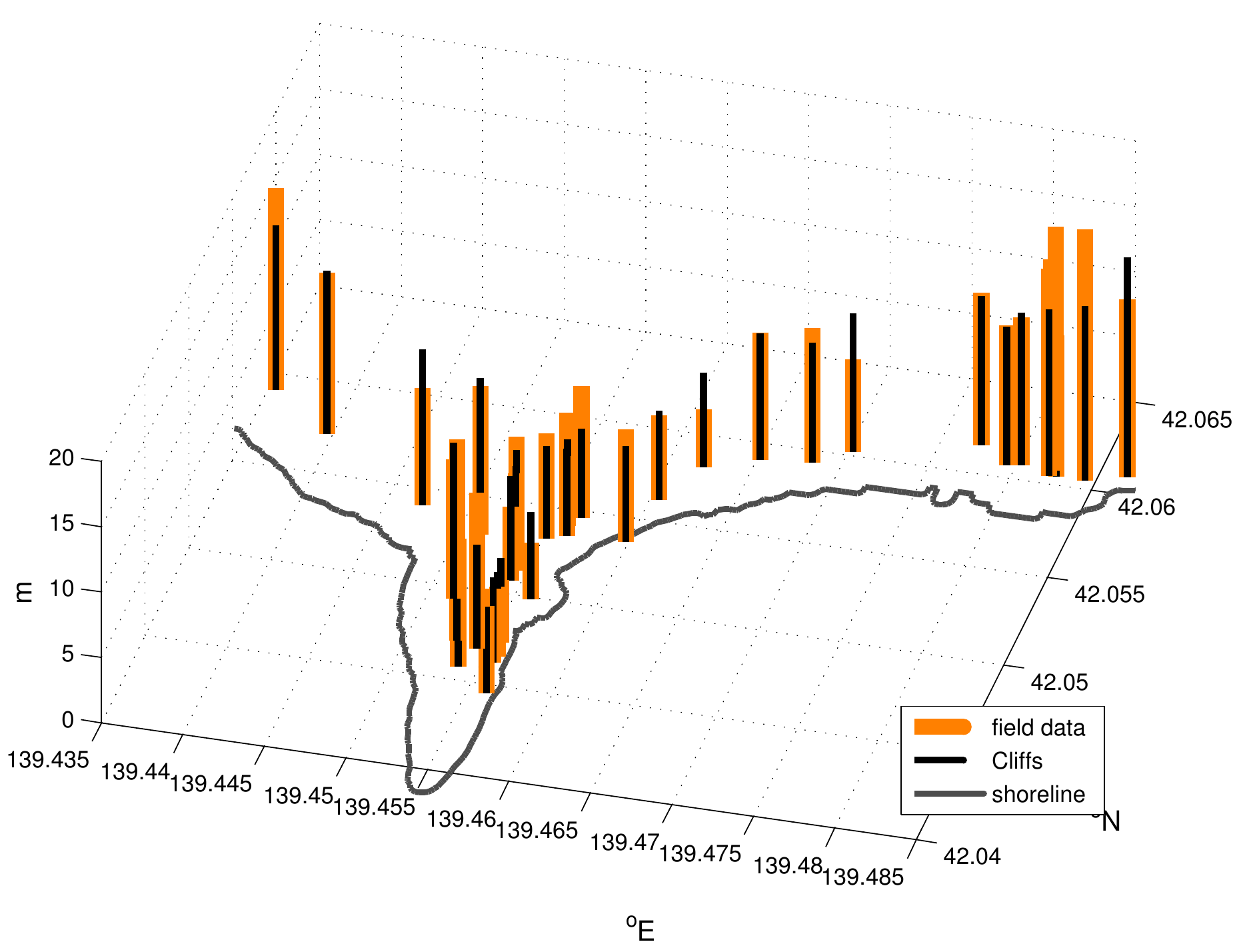} 
	\label{AOrunup}
\end{SCfigure}

\begin{SCfigure}
	\caption{Computed maximal runup height over inundated Monai area.}
		\includegraphics[width=0.5\textwidth]{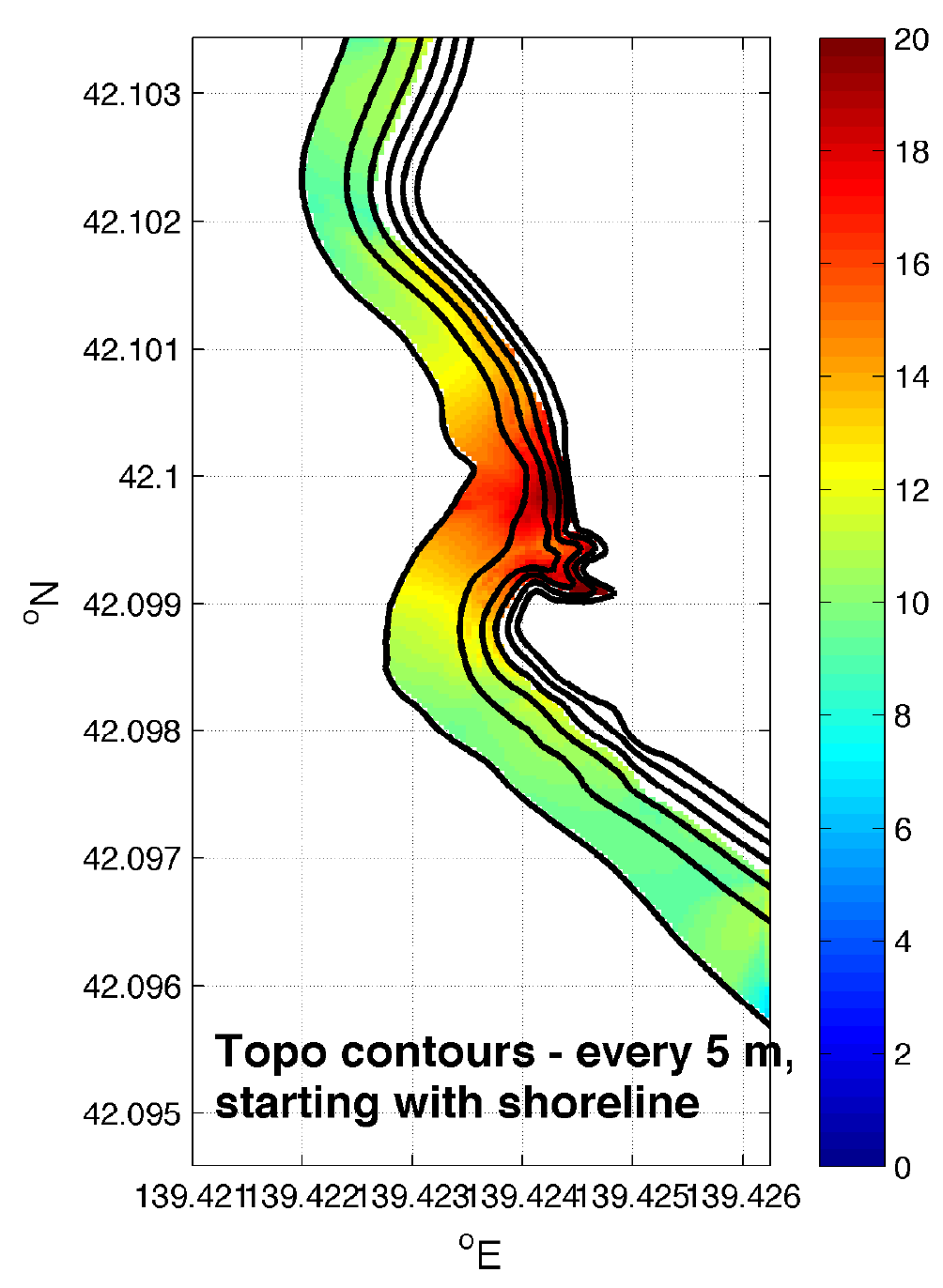} 
	\label{MBrunup}
\end{SCfigure}

Figures \ref{AOrunup} and \ref{MBrunup} display computed maximal runup heights in the high-resolution grids focused on Aonae and Monai areas. Maximum modeled runup height obtained at Hamatsumae was 13 m (reported up to 13.2 m), on Aonae peninsula 12.5 m (reported 12.4 m ), and in Monai valley 21 m (reported up to 31.7 m). It's common among numerical models to under-estimate runup at Monai valley for the reasons discussed in \cite[]{nicolsky2011}. 

\begin{figure}[ht]
\centering
	\resizebox{\textwidth}{!} 
		{\includegraphics{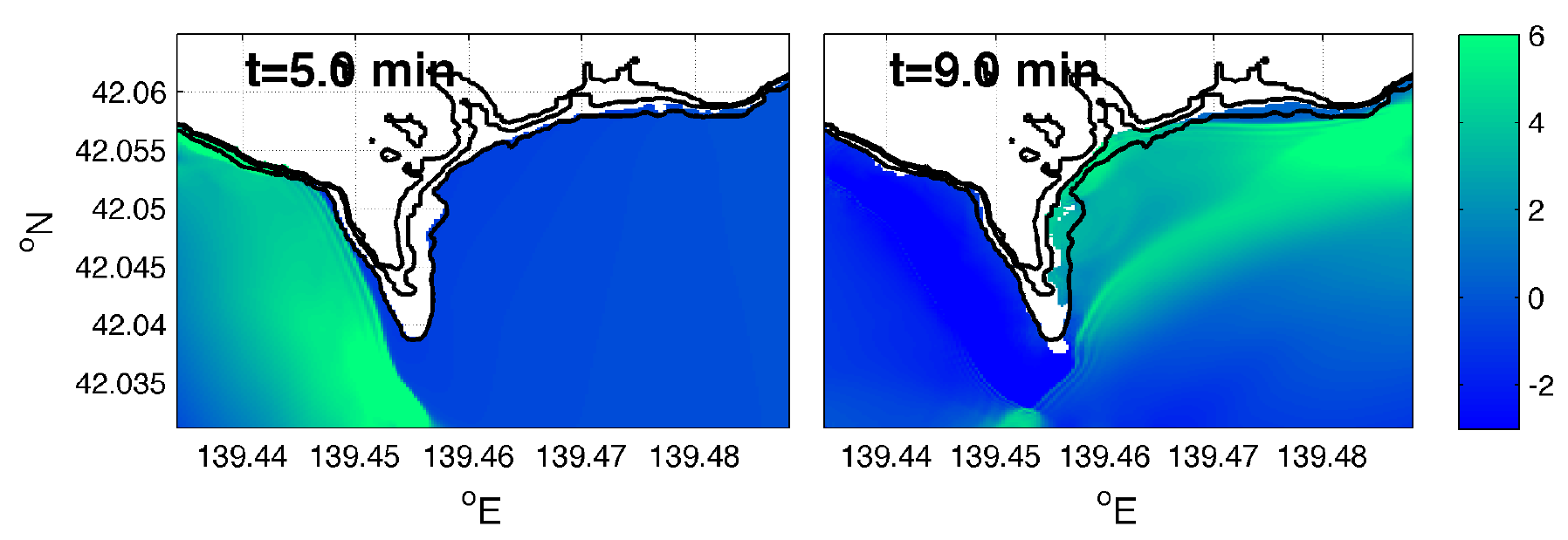}}
	\caption{Snapshots of the simulation in Aonae grid showing the first wave coming from the west 5 min after the EQ, and the second wave coming from the east 9 min after the EQ.}
	\label{waves}
\end{figure}

The second Objective of this BP is to reproduce wave dynamics around Aonae peninsular, namely to simulate arrival of the first wave to Aonae 
5 min after the earthquake coming from the west, and the second wave coming from the east a few mins later. Snapshots of computed sea surface in the Aonae grid 5 and 9 min after the EQ shown in Figure \ref{waves} clearly display the wanted waves.

\begin{SCfigure}
	\caption{Modeled (red) and observed (black circles) sea levels at Iwanai and Esashi tide gages.}
		\includegraphics[width=0.7\textwidth]{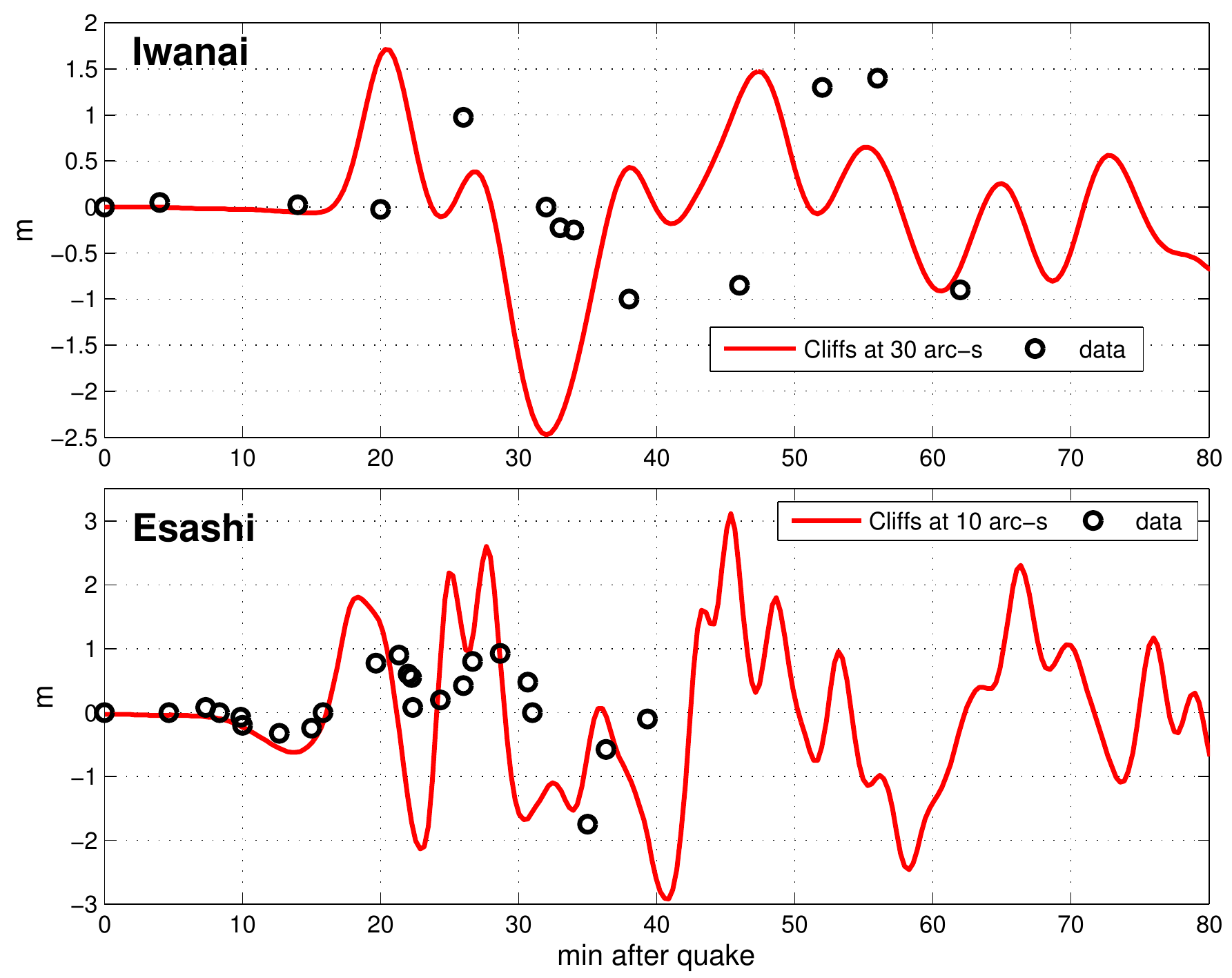} 
	\label{gages}
\end{SCfigure}

Lastly, Figure \ref{gages} presents modeled and observed sea levels at Iwanai and Esashi tide gages.
The comparison between the model and the observations might be little informative, given that available bathymetry at these sites has too coarse resolution for a coastal location (30 arc-s and 10 arc-s), and that observations at Iwanai are scarsely sampled.

\begin{figure}[ht]
\centering
	\resizebox{\textwidth}{!} 
		{\includegraphics{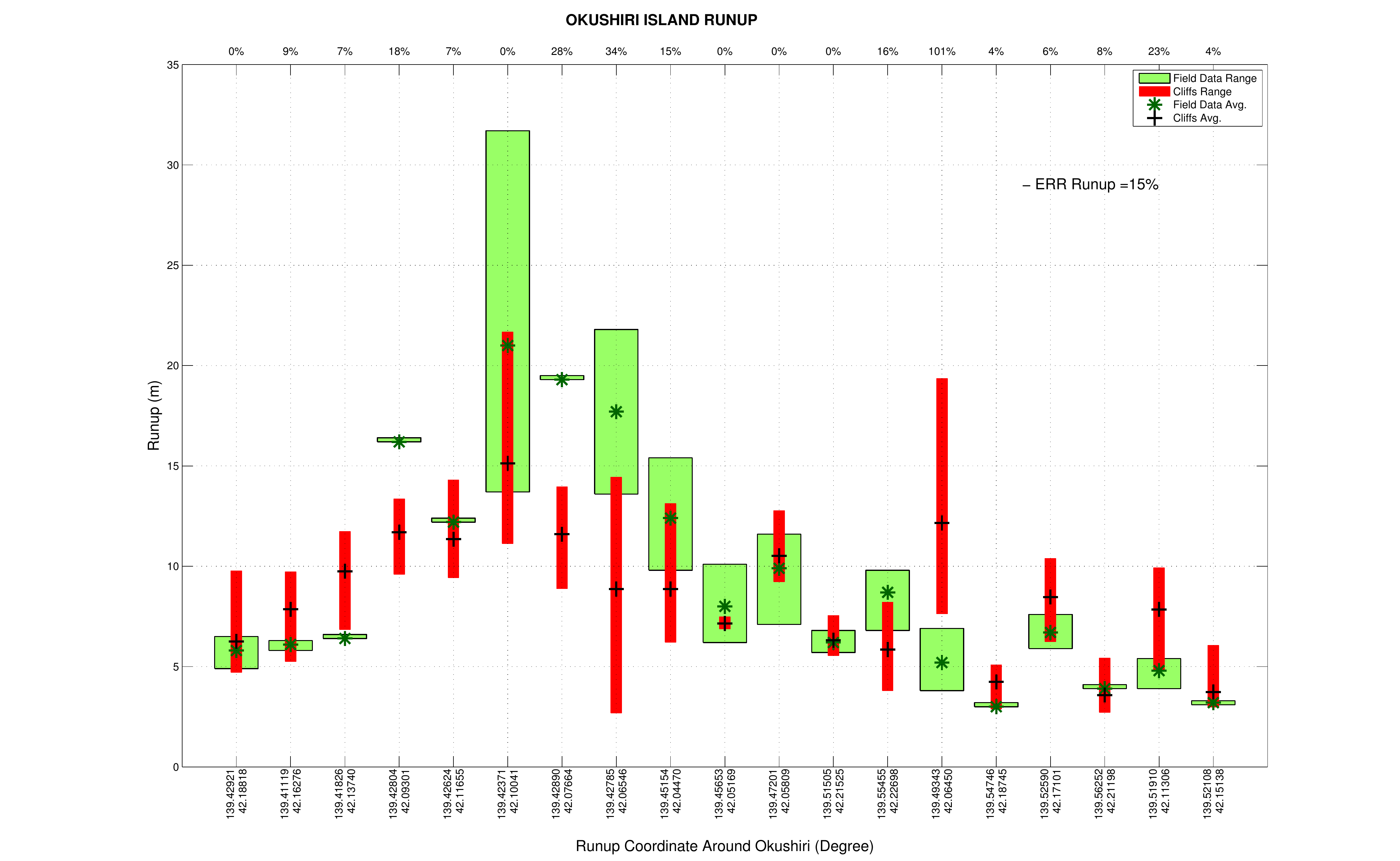}}
	\caption{Runup around Okushiri island vs survey data. Plotted with NTHMP provided script. }
	\label{OKrunup}
\end{figure}

\clearpage
\begin{thebibliography}{99}

\bibitem[Briggs et al.(1995)]{briggs}
Briggs, M. J., Synolakis, C. E., Harkins, G. S., and Green, D. R. (1995),
Laboratory experiments of tsunami runup on a circular island. Pure
Appl. Geophys., 144, 3/4, 569Ð593.

\bibitem[Burwell et al.(2007)]{burwell}
Burwell, D., Tolkova, E., and Chawla, A. (2007), Diffusion and Dispersion Characterization of a Numerical Tsunami Model. Ocean Modelling, 19 (1-2), 10-30. 
doi:10.1016/j.ocemod.2007.05.003

\bibitem[LeVeque(2002)]{leveque2002}
LeVeque, R.J., Finite volume methods for hyperbolic problems. (Cambridge University Press, UK 2002).

\bibitem[Li and Raichlen(2002)]{liraich2002}
Li, Y., and Raichlen, F. (2002), Non-breaking and breaking solitary wave run-up. J. Fluid Mech., 456, 295Ð318.

\bibitem[Liu et al.(1995)]{liu1995}
Liu, P.L.-F., Cho, Y.-S., Briggs, M., Kanoglu, U., and Synolakis, C. (1995), Runup of solitary waves on a circular island. Jornal of Fluid Mech. 302, 259-285.

\bibitem[Liu et al.(2008)]{liu2008}
Liu, P.L.-F., Yeh, H., and Synolakis C. (2008), Advanced Numerical Models for Simulating Tsunami Waves and Runup. Advances in Coastal and Ocean Engineering, 10, 223-230.

\bibitem[Nicolsky et al.(2011)]{nicolsky2011}
Nicolsky, D.J., Suleimani, E.N., and Hansen, R.A. (2011), Validation and verification of a numerical model for tsunami propagation and runup. Pure Appl. Geophys. 168, 1199-1222.

\bibitem[NTHMP(2011)]{nthmp}
[NTHMP] National Tsunami Hazard Mitigation Program, July 2012. Proceedings and Results of the 2011 NTHMP Model Benchmarking Workshop. Boulder: U.S. Department of Commerce/NOAA/NTHMP; NOAA Special Report. 436 p.

\bibitem[Shi et al.(2012)]{shi2012}
Shi, F., Kirby, J.T., Harris, J.C., Geiman, J.D., Grilli, S.T. (2012), A high-order adaptive time-stepping TVD solver for Boussinesq modeling of breaking waves and coastal inundation. Ocean Modelling, 43-44, 36-51.

\bibitem[Stoker(1957)]{stoker}
Stoker, J.J. (1957), Water Waves (Interscience Pub., Inc., New York, NY, USA).

\bibitem[Strang(1968)]{strang}
Strang, G. (1968), On the construction and comparison of difference schemes. SIAM Journal on Numerical Analysis, 5(3), 506-517.

\bibitem[Synolakis(1987)]{synolak1987}
Synolakis, C.E. (1987), The runup of solitary waves. J. Fluid Mech., 185, 523-545.

\bibitem[Synolakis et al.(2007)]{synolak2007}
Synolakis, C.E., Bernard, E.N., Titov, V.V, Kanoglu, U., and Gonzalez, F.I. (2007), Standards, criteria, and procedures for NOAA evaluation of tsunami numerical models. NOAA Tech. Memo. OAR PMEL-135, NTIS: PB2007-109601, NOAA/Pacific Marine Environmental Laboratory, Seattle, WA, 55 pp.

\bibitem[Takahashi(1996)]{takahashi}
Takahashi, T. (1996), Benchmark problem 4: the 1993 Okushiri tsunami - Data, conditions and phenomena. In Long-Wave Runup Models, World Scientific, 384-403.

\bibitem[Titov and Synolakis(1995)]{titov1995}
Titov, V. V., and Synolakis, C. E. (1995), Modeling of breaking and nonbreaking long-wave evolution and runup using VTCS-2, J. Waterw., Port, Coastal, Ocean Eng., 121(6), 308Ð316.

\bibitem[Titov and Gonzalez(1997)]{titov97}
Titov, V., and Gonzalez  F.I. (1997), Implementation and testing of the Method of Splitting Tsunami (MOST) model. NOAA Tech. Memo. ERL PMEL-112 (PB98-122773), NOAA/Pacific Marine Environmental Laboratory, Seattle, WA, 11 pp.

\bibitem[Titov and Synolakis(1998)]{titov1998}
Titov, V. V.,  and Synolakis, C. E. (1998), Numerical modeling of tidal wave runup, J. Waterw., Port, Coastal, Ocean Eng., 124(4), 157Ð171.


\bibitem[Tolkova(2008)]{curvimost}
Tolkova, E. (2008) Curvilinear MOST and its first application: Regional Forecast version 2. In: 
Burwell, D., Tolkova, E. Curvilinear version of the MOST model with application to the coast-wide tsunami forecast, Part II. NOAA Tech. Memo. OAR PMEL-142, 28 pp.
        
\bibitem[Tolkova(2012)]{tolk-nthmp}
Tolkova, E. (2012), MOST (Method of Splitting Tsunamis) Numerical Model. In: [NTHMP] National Tsunami Hazard Mitigation Program. Proceedings and Results of the 2011 NTHMP Model Benchmarking Workshop. Boulder: U.S. Department of Commerce/NOAA/NTHMP; NOAA Special Report. 436 p.

\bibitem[Tolkova(2014)]{cliffs}
Tolkova, E. (2014), LandÐWater Boundary Treatment for a Tsunami Model With Dimensional Splitting. Pure and Applied Geophysics, 171( 9), 2289-2314. 
DOI 10.1007/s00024-014-0825-8

\bibitem[Tolkova(2014a)]{cliffsA}
Tolkova, E. (2014a), Comparative simulations of the 2011 Tohoku tsunami with MOST and Cliffs.
http://arxiv.org/abs/1401.2700

\bibitem[Yanenko(1971)]{yan}
Yanenko N.N. (1971). The method of Fractional Steps. Translated from Russian by M.Holt. Springer, New York, Berlin, Heidelberg. 

\end{thebibliography}
\end{document}